\documentclass[sigplan,screen]{acmart}
\AtBeginDocument{
  }

\setcopyright{none} 
\copyrightyear{Submitted 31 January 2026; accepted 13 February 2026}

\acmConference[HotEthics 2026]{The Workshop on Ethical Systems and Architecture Design}{March 22, 2026}{Pittsburgh, PA, USA}
\acmISBN{}
\acmDOI{}

\usepackage{amsmath,amsfonts}
\usepackage{graphicx}
\usepackage{textcomp}
\usepackage{xcolor}
\usepackage{makecell}
\usepackage{wasysym}

\usepackage{fancyhdr}
\usepackage[normalem]{ulem}
\usepackage{tikz}
\usepackage{pgfplots}
\usetikzlibrary{calc,positioning}
\usepackage{algorithm}
\usepackage[noend]{algorithmic}
\usepackage{soul}
\usepackage{xspace}
\usepackage{enumitem}
\usepackage{physics}
\usepackage{mdframed}

\usepackage{booktabs}

\usepackage{listings}
\setlist{nosep}
\setlist[itemize]{leftmargin=*}
\setlist[enumerate]{leftmargin=*}

\newif\iftight
\tighttrue

\newif\ifcomments
\commentstrue

\newcommand{\ignore}[1]{}
\newcommand{\ie}{\textit{i.e.},\xspace}
\newcommand{\eg}{\textit{e.g.},\xspace}
\newcommand{\bench}[1]{\texttt{\small #1}}
\newcommand{\pgheading}[1]{\noindent\textbf{#1.}}

\newcommand{\titletext}{Challenges and Design Considerations for Finding CUDA Bugs Through GPU-Native Fuzzing}

\title{\titletext}

\author{Mingkai Li}
\affiliation{
  \institution{Columbia University}
  \country{USA}
}
\email{mingkai.li@columbia.edu}

\author{Joseph Devietti}
\affiliation{
  \institution{University of Pennsylvania}
  \country{USA}
}
\email{devietti@cis.upenn.edu}

\author{Suman Jana}
\affiliation{
  \institution{Columbia University}
  \country{USA}
}
\email{suman@cs.columbia.edu}

\author{Tanvir Ahmed Khan}
\affiliation{
  \institution{Columbia University}
  \country{USA}
}
\email{tk3070@columbia.edu}



\begin{document}

\begin{abstract}
Modern computing is shifting from homogeneous CPU-centric systems to heterogeneous systems with closely integrated CPUs and GPUs. %
While the CPU software stack has benefited from decades of memory safety hardening, the GPU software stack %
remains dangerously immature. This discrepancy presents a critical ethical challenge: the world's most advanced AI and scientific workloads are increasingly deployed on vulnerable %
hardware components.

In this paper, we study the key challenges
of ensuring memory safety on heterogeneous systems. We show that, while the number of exploitable bugs in heterogeneous systems rises every year, current mitigation methods often rely on \textit{unfaithful} translations, \ie converting GPU programs to run on CPUs for testing, which fails to capture the architectural differences between CPUs and GPUs. We argue that the faithfulness of the program behavior is at the core of %
secure and reliable heterogeneous systems design.
To ensure faithfulness, we discuss several design considerations of
a GPU-native fuzzing pipeline for CUDA programs.
\end{abstract}

\maketitle

\section{Introduction}
\label{sec:intro}

The end of transistor scaling~\cite{leiserson2020there, dally2020domain, theis2017end} has forced the industry to aggressively shift from homogeneous CPU-centric systems to heterogeneous architectures with tightly coupled CPUs and parallel accelerators, such as GPUs. Despite the fact that these heterogeneous systems are now the backbone of some of the most critical infrastructures, from large-scale machine learning models~\cite{zhao2025insights} in data centers to high-performance scientific simulations~\cite{wahlgren2025dissecting} in supercomputers, they are 
vulnerable in terms of security~\cite{guo2024gpu, zhudemystifying} and reliability~\cite{zheng2025save, lin2025understanding}. 
In particular, while the CPU software stack has benefited from decades of hardening, \eg extensive static~\cite{chess2004static} and dynamic~\cite{liang2018fuzzing} analysis, memory safety tooling~\cite{szekeres2013sok}, and memory safe languages~\cite{matsakis2014rust}, the GPU software stack is comparatively newer and still developing its own testing and debugging ecosystem~\cite{compsan, cugdb}. As a result, prior work~\cite{guo2024gpu, zhudemystifying, zhu2024crossfire} has shown that bugs in heterogeneous systems can lead to sensitive user data leakage, silent data corruptions, or even allow attackers to bypass the host's security hardening mechanisms.

To bridge this gap in heterogeneous systems security, we argue that it is the system designer's ethical responsibility to validate the correctness of GPU programs natively on GPUs. Compromising this principle, current mitigation methods~\cite{singh2026cufuzz, cputriton, santriton} transform heterogeneous GPU programs into homogeneous CPU programs for testing, leading to inaccuracies %
that allow critical bugs to bypass these mitigations.

Therefore, in this paper, we propose a GPU-native fuzzing pipeline that moves the testing logic directly to the hardware component that the GPU program runs on. Specifically, we discuss the challenges and the design considerations of implementing a GPU-native sanitization and fuzzing pipeline on commodity hardware. Utilizing dynamic binary instrumentation, context-sensitive fuzzing, and type-aware mutations, we aim %
to ensure memory safety on heterogeneous systems.

\section{The Ethical Gap in Heterogeneous Systems Security}
\label{sec:analysis}

In this section, we characterize the widening security gap in heterogeneous systems due to GPU memory safety bugs and analyze why existing mitigation methods fail to ensure faithful protection on heterogeneous systems.

\begin{figure}[t]
\centering
\small
\includegraphics[width=\linewidth]{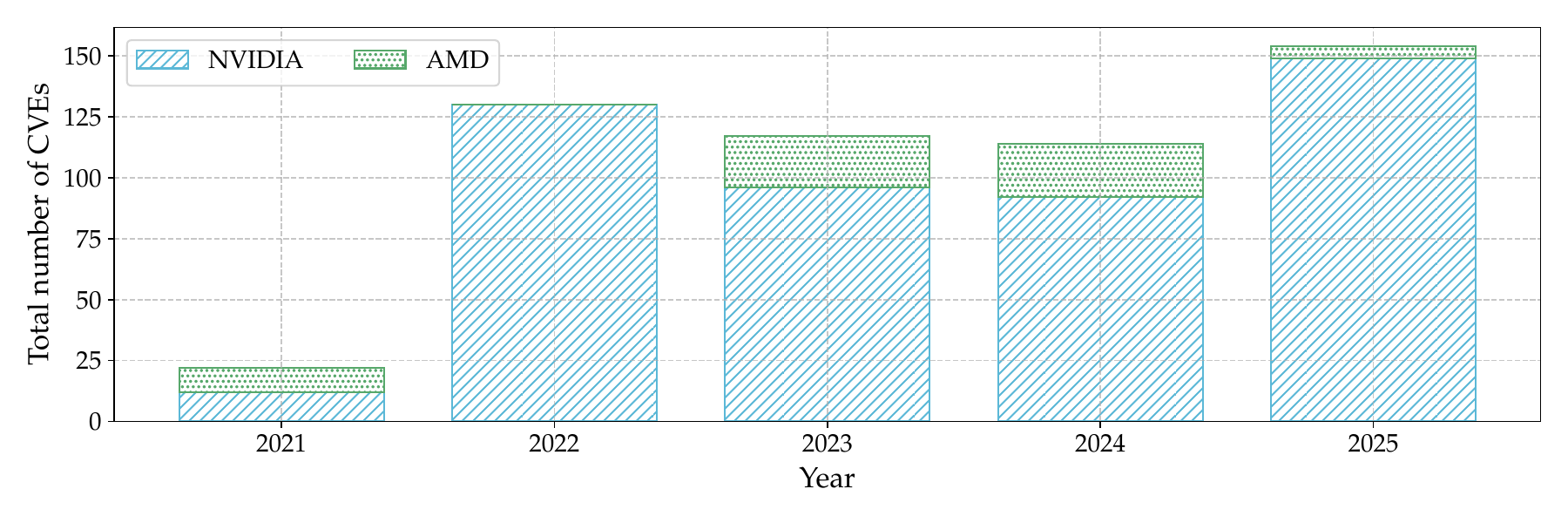}
\caption{As heterogeneous systems become popular, the number of exploitable bugs in these systems also rises.}
\label{fig:cve-counts-per-year}
\end{figure}

\pgheading{Rising numbers of GPU bugs} We investigate the prevalence of GPU bugs by counting the number of exploitable bugs~\cite{nvsec, amdsec}, \ie Common Vulnerabilities and Exposures (CVE), over the years. Figure~\ref{fig:cve-counts-per-year} shows the corresponding results for two major GPU vendors: (1) NVIDIA and (2) AMD. As we show, with the rising popularity of machine learning workloads, the number of exploitable GPU bugs are rapidly rising over the last few years.

\pgheading{Implications of GPU memory safety bugs} GPU memory safety bugs have significant implications. In particular, they lead to return-oriented programming (ROP) attacks~\cite{guo2024gpu, roels2025cuda}. Such attacks undermine the effectiveness of machine learning workloads~\cite{guo2024gpu}, and even allow attackers to perform arbitrary computations with a Turing-complete gadget set~\cite{roels2025cuda}. Furthermore, recent work~\cite{sorensen2024leftoverlocals} also shows that malicious attackers can exploit these bugs to compromise the confidentiality of machine learning workloads.

\pgheading{Existing work and why they fall short} Recognizing the urgent need of GPU memory safety, academic researchers and industry practitioners have proposed a wide range of techniques~\cite{singh2026cufuzz,zhou2025fuzz4cuda, lu2026cusafe, cucatch-pldi23, compsan, erb2018clarmor, di2018gmod, lee2025let, gpu-tag-memory-isca23, ziad2025gpuarmor} 
to detect GPU memory safety bugs. Unfortunately, these techniques fail to achieve faithfulness, efficiency, and scalability at the same time. %
Specifically, existing work~\cite{singh2026cufuzz, cputriton, santriton} transforms heterogeneous GPU programs into homogeneous CPU programs to leverage existing CPU-focused infrastructure~\cite{fioraldi2020afl++, asan}. While such solutions help detect bugs in GPU programs, the complex architectural differences between CPUs and GPUs make it very challenging to ensure the faithfulness of the program behavior across this translation process, causing false positives and negatives. Moreover, existing GPU-native solutions~\cite{zhou2025fuzz4cuda, compsan, lu2026cusafe, cucatch-pldi23} %
either incurs high run-time overhead~\cite{compsan}, low fuzzing throughput~\cite{zhou2025fuzz4cuda}, poor detection accuracy~\cite{di2018gmod, erb2018clarmor}, or can not be applied to many real-world GPU applications~\cite{lu2026cusafe, cucatch-pldi23} due to NVIDIA's closed-source ecosystem~\cite{nvlibrary, nvcc, cutlass}. %
Furthermore, some existing solutions~\cite{lee2025let, gpu-tag-memory-isca23, ziad2025gpuarmor} require customized hardware changes, and thus are not deployable on commodity GPUs. %
Given these limitations, it is significantly challenging to detect GPU memory safety bugs through fuzzing, a widely-used technique to detect CPU memory safety bugs. Next, we describe the key challenges of adopting fuzzing for GPUs in \S\ref{sec:motivation}.

\section{Challenges of 
Securing
CUDA Programs}
\label{sec:motivation}

In this section, we discuss the key
challenges of ensuring memory safety for CUDA programs through fuzzing: (1) lack of sanitization, (2) input mutation, (3) coverage tracking, and (4) fuzzing harnesses.

\pgheading{Lack of address sanitization} Address sanitization is a widely-used dynamic technique to detect memory safety bugs that are difficult to find using static analysis~\cite{chess2004static}. Address sanitization enables finding such bugs by adding a number of dynamic checks or instrumentation~\cite{asan}. Such an instrumentation accesses metadata to detect bugs like buffer overflows and use-after-frees~\cite{cucatch-pldi23}. Once the instrumentation detects such a bug, it crashes the program's execution, improving the bug detection capability of automated techniques like fuzzing~\cite{liang2018fuzzing}. While there exists a wide range of address sanitization techniques for CPUs~\cite{asan, asan--, duck2018effectivesan, giantsan, zhang2021sanrazor, yu2024shadowbound, nagarakatte2009softbound, hwasan},
they do not generalize to GPUs~\cite{cucatch-pldi23,lu2026cusafe}.

\pgheading{Lack of input mutation} Automated techniques like fuzzing start testing programs with an initial set of user-provided input seeds~\cite{pailoor2018moonshine} and randomly change them to detect bugs~\cite{godefroid2020fuzzing}. As randomly generated inputs are mostly invalid~\cite{godefroid2020fuzzing}, \ie programs reject them quickly, fuzzing techniques improve the possibility to generate valid inputs through input mutation~\cite{ivysyn}. Input mutation makes small changes to existing inputs that may still keep the input valid, while also testing new program behavior, such as application logic not tested previously. While there exist many input mutation techniques to detect CPU memory safety bugs~\cite{godefroid2020fuzzing}, they unfortunately do not apply to GPUs~\cite{singh2026cufuzz,zhou2025fuzz4cuda}. This gap exists because current fuzzing frameworks lack the GPU architecture–specific domain knowledge required to design mutation operators that can effectively trigger memory safety bugs in highly parallel CUDA kernels.

\pgheading{Lack of coverage tracking} Automated bug detection techniques, such as fuzzing, aim to test a program across a large number of unique behavior or application logic~\cite{shah2022mc2}. To quantify the effectiveness of testing different application logic, fuzzing techniques leverage code coverage that measures what fraction of a program gets executed during a test for a specific input~\cite{bohme2016coverage}. If a specific input triggers executions of new control-flow edges (\ie branches), or new statements in general, fuzzing techniques also reuse the input for further mutation~\cite{godefroid2020fuzzing}. As code coverage enables fuzzing techniques to measure test effectiveness, while also helping generate interesting inputs, coverage-guided fuzzing has become an effective technique to find bugs in real-world applications~\cite{serebryany2016continuous,bohme2016coverage,rawat2017vuzzer}. Alas, coverage-guided fuzzing is challenging for heterogeneous systems due to difficulty to track coverage for CUDA code that run on GPUs~\cite{singh2026cufuzz,zhou2025fuzz4cuda}.

\pgheading{Lack of fuzzing harness} To test programs for memory safety bugs, fuzzing techniques also require fuzzing harnesses to invoke programs with mutated inputs~\cite{lemieux2018fairfuzz}. Fuzzing harness ensures testing programs properly by setting up specific contexts programs require~\cite{sherman2025no}. Initializing programs with necessary contexts, fuzzing harness avoids costly false positives or negatives~\cite{sherman2025no}, increasing the utilization of testing resources~\cite{sherman2025no}. Unfortunately, existing fuzzing harnesses to test CPU programs do not generalize to GPUs as CUDA programs running on GPUs require specific context initialization, just-in-time (JIT) compilation, and CPU-GPU collaborative execution~\cite{singh2026cufuzz,zhou2025fuzz4cuda}.

\section{The GPU-Native Design}
\label{sec:design}

To address these challenges, we propose a GPU-native design that utilizes dynamic binary instrumentation to faithfully sanitize and fuzz CUDA programs.

\subsection{GPU-Native Address Sanitization and Coverage Tracking}
\label{subsec:sanitization}

\pgheading{Address Sanitizer} We are designing an address sanitization software tool to detect memory safety bugs for GPUs. A key requirement for an address sanitizer to be practical is to accurately detect memory safety bugs for as many programs as possible at a reasonable overhead. We are achieving this by performing the address sanitization natively on the GPUs instead of CPUs.

We are building an address sanitizer that can detect memory safety bugs in both closed-source and open-source CUDA kernels on off-the-shelf NVIDIA GPUs. Unlike existing solutions that require custom hardware~\cite{lee2025let} or support only open-source CUDA kernels~\cite{lu2026cusafe}, our address sanitizer will detect memory safety bugs even for closed-source CUDA kernels in commodity NVIDIA GPUs. We are implementing such an address sanitizer using NVBit~\cite{villa2019nvbit}, NVIDIA's dynamic binary instrumentation tool. Using NVBit, we are performing the address sanitization on the GPU itself. The GPU's parallelism will help our implementation speed up address sanitization. Such a GPU-native solution will also enable our technique to detect new types of GPU memory safety bugs that existing techniques fail to detect.

Using NVBit, our sanitizer instruments GPU memory access instructions to detect memory safety bugs. Our sanitizer performs this bug detection along with the execution of GPU CUDA kernels by leveraging GPU parallelism. In particular, our sanitizer will maintain metadata for each unit (\eg 4 bytes) of GPU global, local, and shared memory, along with metadata for pointers. While accessing memory bytes with a pointer, %
our sanitizer looks up the metadata corresponding to both the memory bytes and pointers to identify different classes of GPU memory safety bugs.

\pgheading{Coverage Tracking} We are also building a software-only coverage profiling mechanism. The goal of our coverage profiler is to help detect memory safety bugs by improving the effectiveness of fuzzing. Toward this goal, our coverage profiler supports both open-source and closed-source CUDA kernels. To help detect as many bugs as possible, our coverage profiler works on commodity NVIDIA GPUs. Similar to our address sanitizer, our coverage profiler also achieves these goals by using NVBit.

Using NVBit, our coverage profiler instruments GPU control flow instructions during the execution of CUDA kernels. Specifically, our coverage profiler maintains metadata for each control flow instruction. While executing a control flow instruction, the instrumentation logic of our coverage profiler updates the corresponding metadata to count the number of executions. We use these execution counts as feedback to improve the effectiveness of our fuzzer (\S\ref{subsec:fuzzing}).

\subsection{Context-Sensitive Fuzzing and Type-Aware Mutations}
\label{subsec:fuzzing}

Detecting memory safety bugs on GPUs requires invoking GPU CUDA kernels with specific contexts and inputs.
To ensure specific contexts and inputs, we are proposing context-specific fuzzing and type-aware mutations.

\pgheading{Context-Sensitive Fuzzing} Closed-source NVIDIA libraries often involve multiple layers of abstraction without exposing many low-level kernels for direct usage. Instead, these low-level kernels can only be invoked by high-level kernels. As a result, testing a significant fraction of these libraries' functionality requires setting up a specific chain of functions active on the call stack, \ie \textit{context}. Furthermore, while running CUDA kernels, GPUs compile the machine-independent Parallel Thread Execution (PTX) code to machine-dependent Source and Assembly (SASS) code just in time~\cite{cucatch-pldi23}. Just-in-time compilation in such a specific context also incurs significant overhead for fuzzing~\cite{zhou2025fuzz4cuda}.
Therefore, we are enabling context-sensitive fuzzing with low overhead by leveraging the open-source CUDA library samples~\cite{cuda-library-samples-github}.

The open-source CUDA library samples contain extensive examples that NVIDIA provides to demonstrate the usage of its closed-source CUDA libraries. These examples invoke high-level kernels by setting up the necessary contexts, while also running low-level kernels. We are leveraging these library examples to ensure proper contexts for fuzzing. To use these examples for fuzzing, we will divide their execution among different phases, such as \textit{initialization} phase, \textit{computation} phase, \textit{termination} phase.

The initialization phase will include calling functions to set up the contexts, allocating memory, and copying memory bytes from CPUs to GPUs. The computation phase will include calling the higher-level kernel functions to start GPU computations, while also executing following lower-level kernels. Finally, the termination phase includes calling functions to copy memory bytes from GPUs to CPUs, while also synchronizing and freeing memory bytes.

To divide the execution of library examples among these different phases, we will leverage compiler-based automated techniques~\cite{wu2016gpucc, jeong2025inside}
while also exploring LLM-assisted manual strategies~\cite{ouyang2025kernelbench, lange2025ai}.
Once we divide the execution among different phases, we will amortize the initialization and termination phases across many instances of the computation phase. In particular, we will wrap the execution of the computation phase with our own fuzzing loop. This fuzzing loop will record the input parameters to the high-level kernels, while also setting up the coverage profiling and sanitization infrastructure. Then, the fuzzing loop will run one instance of the computation phase. Based on this instance's outcome (\eg coverage), the fuzzing loop will then mutate the input and rerun the computation phase. Once the fuzzing loop finds an interesting instance of the computation phase (\eg crash, sanitization failure) or completes a pre-determined number of instances for the computation phase, it will finish its execution and invoke the termination phase. By amortizing the initialization and termination phases, our context-sensitive fuzzing will improve the effectiveness and efficiency of memory safety bug detection for GPU CUDA kernels.

\pgheading{Type-Aware Mutations} We are using mutation-based fuzzing to detect memory safety bugs for GPU kernels. In particular, we are leveraging type-aware mutations corresponding to different argument types for different GPU kernels, including integers, floating points, and arrays. Prior work~\cite{ivysyn} has shown that coverage-guided byte-level mutations fail to bypass the shallow argument-type checks for machine learning kernels. %
To bypass such 
checks and test interesting functionality, we are working on type-aware mutations.

As we are currently studying CVEs corresponding to CUDA kernels, we observe that these kernels suffer crashes for some specific input arguments. Such input arguments trigger edge-case behavior, including integers with large positive or negative values overflowing arrays, empty arrays, and arrays with multiple dimensions. Based on our observation, we are using the following mutation strategies for various input arguments for GPU CUDA kernels:

\begin{itemize}
    \item \textbf{Integer arguments}: If a CUDA kernel takes integers as input arguments, we mutate the value of these integer arguments across zero, the maximum positive, and the minimum negative values.
    \item \textbf{Floating point arguments}: Floating point values include different components, such as sign, mantissa, and exponent. Consequently, when a CUDA kernel takes floating-point values as input arguments, we mutate the value of these arguments by mutating these components using several strategies. For example, we vary the 1-bit sign across 0 and 1. For mantissa and exponent, we use several mutation operators, such as flipping one or multiple bits and adding or subtracting values.
    \item \textbf{Arrays}: If a GPU kernel takes arrays as input arguments, we mutate the value of these arguments in two different ways: (1) value mutation and (2) pointer mutation. While mutating array values, we use arrays with large positive and negative values, arrays with dimensions different from the dimension the kernel expects (\eg 2-dimensional array when the kernel expects a 1-dimensional array). While mutating array pointers, we use pointers to GPU memory space that are different from the GPU memory space the kernel expects. For example, if the GPU kernel expects an argument array allocated on global memory, we mutate it to have a different array allocated on local/shared memory.
\end{itemize}

\section{Preliminary Experimental Results}
\label{sec:eval}

In this section, we demonstrate the capability of our coverage profiler (\S\ref{subsec:sanitization}) in analyzing closed-source CUDA kernels. In particular, we present preliminary experimental results for 11 cuBLAS libraries from NVIDIA's CUDA library samples~\cite{cuda-library-samples-github}.

\pgheading{Experiment Setup} We conduct preliminary experiments on a Linux server running Ubuntu 22.04 and Linux kernel version 5.15.0. The server has two AMD EPYC 7763 processors, 256 GB RAM, and two NVIDIA A100 GPUs. We use NVIDIA driver version 590.48.01 and CUDA version 13.1.

\begin{table}[t]
\centering
\small
\caption{Coverage statistics of 11 cuBLAS library samples. BB in the table indicates basic blocks in GPU kernels. The geometric mean of basic block coverage is only 25.98\%.}
\label{tab:kernel_coverage}
\begin{tabular}{lcccc}
\toprule
\textbf{Library} & \textbf{Total BBs} & \textbf{Hit BBs} & \textbf{BB Cov. (\%)} & \textbf{Hit Edges} \\
\midrule
amax  & 95  & 47  & 49.47 & 61  \\
amin  & 106 & 47  & 44.34 & 59  \\
asum  & 14  & 9   & 64.29 & 13  \\
axpy  & 30  & 7   & 23.33 & 7   \\
copy  & 35  & 6   & 17.14 & 6   \\
dot   & 57  & 23  & 40.35 & 27  \\
nrm2  & 340 & 177 & 52.06 & 189 \\
rot   & 81  & 10  & 12.35 & 10  \\
rotm  & 132 & 12  & 9.09  & 12  \\
scal  & 19  & 4   & 21.05 & 4   \\
swap  & 77  & 10  & 12.99 & 10  \\
\midrule
\textbf{GeoMean} & \textbf{---} & \textbf{---} & \textbf{25.98} & \textbf{---} \\\bottomrule
\end{tabular}
\iftight\vspace{-0.15in}\fi
\end{table}

\pgheading{Results} Table~\ref{tab:kernel_coverage} shows the coverage statistics of 11 library samples from cuBLAS~\cite{cublas}, which is NVIDIA's proprietary, closed-source library that provides highly-optimized GPU-accelerated implementations of fundamental linear algebra operations. 
The results indicate that inputs from CUDA library samples~\cite{cuda-library-samples-github} achieve relatively low code coverage, with a geometric mean of only 25.98\% across the tested library samples. Specifically, while \bench{asum} reaches the highest coverage of 64.29\%, \bench{rotm} exhibits the lowest coverage of merely 9.09\%. This significant gap suggests a substantial 
gap
in state exploration for complex GPU programs, a challenge we aim to address with our GPU-native fuzzing pipeline.

\section{Conclusion}
\label{sec:conclusion}
The widening security gap in heterogeneous systems presents a critical ethical concern. %
Therefore, in this paper, we propose a GPU-native fuzzing pipeline that combines dynamic binary instrumentation with context-sensitive fuzzing to ensure faithful bug detection of CUDA programs. Our design aims to replace existing transformation-based mitigation solutions to achieve more precise bug detection on heterogeneous systems.

\section*{Acknowledgments}
\label{sec:acknowledgment}

This work was supported in part by the Center for Ubiquitous Connectivity (CUbiC) and is sponsored by Semiconductor Research Corporation (SRC) and Defense Advanced Research Projects Agency (DARPA) under the JUMP 2.0 program. Results presented in this paper were obtained using Chameleon Cloud~\cite {chameleon} supported by the National Science Foundation.

\bibliographystyle{ACM-Reference-Format}
\bibliography{refs}

\end{document}